\documentclass[fleqn,twoside]{article}
\usepackage{espcrc2}
\usepackage{amsmath}
\usepackage{slashed}
\usepackage[numbers,sort&compress]{natbib}

\title{\vspace{-5cm}
\hfill {\normalsize HU-EP-10/32}\\
\hfill {\normalsize DESY 10-089}\\
\hfill {\normalsize SFB/CPP-10-52}\vspace{1cm}\\
Analytic computations of massive one-loop amplitudes}

\author{Simon Badger\address[DESYz]{Deutches Elektronen-Synchrotron DESY, Platanenallee 6, D-15738 Zeuthen,
Germany}\thanks{Talk presented at Loop and Legs in Quantum Field Theory, W\"{o}rlitz, Germany, 25th-30th April 2010},
	Ralf Sattler\address[HU]{ Institut f\"ur Physik, Humboldt-Universit\"at zu Berlin,
	Newtonstra\ss{}e 15, D-12489 Berlin, Germany},
	Valery Yundin\addressmark[DESYz]}

\def\fl#1{ {#1}^\flat}
\def\flm#1{ {#1}^{\flat,\mu}}
\def\tb{{\bar{t}}}
\def\e{\epsilon}
\def\la{\langle}
\def\ra{\rangle}

\def\mc{\mathcal}

\def\A#1#2{\la#1#2\ra}
\def\B#1#2{[#1#2]}
\def\AB#1#2#3{\la#1|#2|#3]}

\begin{document}

\begin{abstract}
  We show some new applications of on-shell methods to calculate compact helicity amplitudes for $t\tb$ production
  through gluon fusion. The rational and mass renormalisation contributions are extracted from two independent Feynman diagram based approaches.
\end{abstract}

\maketitle

\section{Introduction}

With the first data from the LHC now arriving the need for full NLO studies of multi-leg QCD processes is becoming ever
more important. Theoretical developments in recent years have rapidly opened the range of processes that can be
feasibly considered. Modern methods using combinations of unitarity, factorisation and algebraic tensor reduction have
been able to produce automated numerical codes for phenomenological studies of $2\to 4$ processes
\cite{Berger:2008sj,Berger:2009ep,Berger:2009zg,Ellis:2009zw,Ellis:2009bu,Bevilacqua:2009zn,Bevilacqua:2010ve,Bredenstein:2009aj}.
A more detailed overview of the current status of NLO studies can be found in ref. \cite{Binoth:2010ra}.

One of the most important areas for both phenomenological applications and as a benchmark for theoretical complexity is
the inclusion of full mass dependence in NLO QCD studies. Numerical methods have been at the forefront of progress in
this direction and impressive results have already been achieved for $pp\to
t\tb+j$~\cite{Dittmaier:2007wz,Bevilacqua:2010ve,Melnikov:2010iu}, $pp\to
t\tb+b\bar{b}$~\cite{Bredenstein:2009aj,Bredenstein:2010rs,Bevilacqua:2009zn} and more recently $pp\to
t\tb+2j$~\cite{Bevilacqua:2010ve}.

Progress in fully analytic computations of massive one-loop amplitudes has been relatively slow. Despite many
theoretical developments \cite{Bern:1995db,Britto:2006fc,Britto:2008vq,Britto:2009wz,Kilgore:2007qr}, there are
relatively few complete
calculations used in phenomenological studies. The hope that analytical computations would provide faster
and numerically more stable results motivates us to study this problem in more detail and investigate to what extent the
problem can be automated. In this contribution we present new compact analytic one-loop amplitudes for $gg\to t\tb$
scattering. In section \ref{sec:gu} we briefly outline the on-shell techniques we have used and discuss the problem
of mass-renormalisation within this approach. In section \ref{sec:feynman} we discuss two independent Feynman
calculations which have been used to verify our results and generate compact representations of the rational terms.
We define our conventions for the Spinor-Helicity formalism in section \ref{sec:spinor} before presenting our results
in \ref{sec:ttgg}. We discuss some possibilities for alternative methods for determining tadpole coefficients before
reaching our conclusions.

\section{Generalised Unitarity \label{sec:gu}}

Generalised unitarity has become an essential tool in the computation of multi-particle one-loop amplitudes over the
last few years. Building from the pioneering work of Bern, Dixon, Dunbar and Kosower in the mid-nineties
\cite{Bern:1994zx} the modern
picture of unitarity gives a purely algebraic approach to the computations of one-loop amplitudes making use of complex
analysis \cite{Britto:2004nc,Ossola:2006us,Forde:2007mi,Ellis:2007br,Giele:2008ve}. 

Quadruple cuts with complex momenta completely freeze the four dimensional loop integration leading directly to the
algebraic evaluation of the associated scalar box integral coefficient \cite{Britto:2004nc}. The technique of integrand
reduction first proposed by Ossola, Papadopoulos and Pittau \cite{Ossola:2006us} was shown to apply elegantly to the
extraction of triangle and bubble coefficients using Laurent expansions of the unconstrained integrations by Forde
\cite{Forde:2007mi}. This method has been generalised to the case of massive amplitudes \cite{Kilgore:2007qr} and is the
basis for the method used here.

In these proceedings we apply the technique of generalised unitarity to compute compact analytic expression for one loop
$t\tb gg$ helicity amplitudes. The results are obtained from a semi-automated system using {\tt FORM}~\cite{Vermaseren:2000nd} and {\tt Maple}.
The integral basis and integrands are generated with {\tt Maple} before the tree amplitudes (from BCFW recursion
relations \cite{Britto:2004ap,Britto:2005fq}) are expanded using {\tt FORM} to extract closed analytic forms for the integral coefficients. IR consistency
equations have been verified and used to constrain the form of the amplitude and ensure a compact representation.

The colour ordered amplitudes can be expressed in $D=4-2\e$ dimensions as a product of rational coefficients multiplied
by divergent scalar integrals,
\begin{align}
  &A^{(1)} = 
  \sum_{K_4} C_{4;K_4}I_{4;K_4}
  + \sum_{K_3} C_{3;K_3}I_{3;K_3}\nonumber\\&
  + \sum_{K_2} C_{2;K_2}I_{2;K_2}
  + C_{1}I_{1}
  + R^{DD}+\mathcal{O}(\e).
\end{align}
In the massive amplitudes the wave-function renormalisation and tadpole contributions are
non-zero and require mass-renormalisation to be performed before the terms proportional to $\log(m^2)$ can be
determined. This presents a problem when formulating a purely on-shell construction of the amplitude since cuts in the
$\log(m^2)$ channels contain explicit divergences. One can circumvent this problem by explicitly removing the
singularities but this causes gauge invariance to be broken in intermediate stages of the computation \cite{Ellis:2008ir}. 
In $4-2\e$ dimensions the universal pole structure is sufficient to determine the $\log(m^2)$ terms however additional rational terms are
introduced,
\begin{align}
  &A^{(1)} = 
  \sum_{K_4} C_{4;K_4}I_{4;K_4}
  + \sum_{K_3} C_{3;K_3}I_{3;K_3}\nonumber\\&
  + \sum_{K_2'} C_{2;K_2'}F_{2;K_2'}
  + C_{2;m^2} I_2(m^2,0,m^2)\nonumber\\&
  + \mathcal{R}+\mathcal{O}(\e).
  \label{eq:a1basis2}
\end{align}
where the sum over $K_2'$ includes only the ``off-shell" bubbles with well defined unitarity cuts.
In our approach both $C_{2;m^2}$ and $\mathcal{R}$ are computed using Feynman diagram based techniques. We can then
verify that the coefficient of $C_{2;m^2}$ matches that predicted by the IR constraints. We then demonstrate for the example
of the $t^+g^+g^+\tb^+$ amplitude that $R^{DD}$ can also be determined using the cut-constructible information. 

\section{Feynman diagram approach \label{sec:feynman}}

The amplitude was computed using two independent approaches. An analytic computation using traditional Passarino-Veltman
tensor reduction was performed with the help of {\tt DIANA}~\cite{Tentyukov:1999is} (diagram generation) and {\tt FORM}~\cite{Vermaseren:2000nd}.
A second method employed a numerical implementation of tensor reduction based on recurrence relations in shifted
dimensions \cite{Davydychev:1991va,Fleischer:1999hq,Fleischer:2010mq,Diakonidis:2008ij}.

The 37 colour-ordered one-loop diagrams contributing to the process $gg\to t\tb$, including
ghosts and on-shell mass renormalisation counter terms, were further processed analytically in two independent {\tt
FORM} codes to generate tensor integral representations of the amplitudes. The calculations were performed using the
spinor-helicity formalism in the Four-Dimensional Helicity(FDH) regularisation scheme.

Full agreement between the unitarity and Feynman approaches was achieved with the use of two independent C++
implementations of the spinor products and the {\tt qcdloop} package for finite parts of the $4-2\e$-dimensional scalar integrals
\cite{Ellis:2007qk}.

\section{Spinor/Helicity Formalism \label{sec:spinor}}

For massless particles it is possible to completely decompose all momenta into a basis of
two-component Weyl spinors since,
\begin{equation}
	p^\mu = \frac{1}{2}\AB{p}{\gamma^\mu}{p}.
	\label{eq:mdecomp}
\end{equation}
The polarisation vectors and fermion wave-functions then fit easily into a helicity basis,
\begin{align}
	u_+(p) &= |p\ra & u_-(p) &= |p] \\
	\e_+^\mu(p,\xi) &= \frac{\AB{\xi}{\gamma^\mu}{p}}{\sqrt{2}\A{\xi}{p}} &
	\e_-^\mu(p,\xi) &= \frac{\AB{p}{\gamma^\mu}{\xi}}{\sqrt{2}\B{p}{\xi}}
\end{align}
Kleiss and Stirling described how to construct
well defined helicity states for massive momenta through the introduction of an arbitrary massless vector which defines
the reference frame \cite{Kleiss:1985yh}. For a massive vector $P$ we use a massless vector $\eta_P$ to define a massless projected vector
$\fl P$:
\begin{equation}
	P^\mu = \flm P + \frac{m^2}{2P\cdot\eta}\eta_P^\mu
\end{equation}
The $\bar{u}$ and $v$ spinors can then be defined by:
\begin{align}
	\bar{u}_\pm(P,m;\fl P,\eta_P) &= \frac{\la\eta_P\mp|(\slashed{P}+m)}{\A{\eta_P\mp|}{\fl P\pm}} \\
	v_\pm(P,m;\fl P,\eta_P) &= \frac{(\slashed{P}-m)|\eta_P\pm\ra}{\A{\fl P\mp|}{\eta_P\pm}}.
\end{align}
The freedom to keep $\eta_P$ arbitrary gives us the ability to relate the positive and negative helicity states by
applying a simple transformation ($\fl P \leftrightarrow \eta_P$):
\begin{equation}
	v_-(P,m;\fl P,\eta_P) = \frac{\A{\fl P}{\eta_P}}{m} v_+(P,m;\eta_P,\fl P).
	\label{eq:heltrans}
\end{equation}
It is therefore sufficient to consider helicity amplitudes where all massive fermion have positive helicity and obtain
the remaining amplitudes by applying eq.(\ref{eq:heltrans}). For the four-point $t\tb gg$ example considered here this
gives us two independent configurations, $++++$ and $++-+$.

\section{$t\bar{t}gg$ helicity amplitudes \label{sec:ttgg}}

In this section we demonstrate the techniques by applying them to the well studied case of top pair
production through gluon fusion. The final results are obtained in an extremely compact analytic form and are in
agreement with the results in the literature \cite{Korner:2002hy,Anastasiou:2008vd,Badger:2008za}. The amplitude is
decomposed into colour ordered primitive amplitudes as outlined in \cite{Bern:1994fz}:
\begin{align}
	&\mc A^{(0)}_4(1_t,2,3,4_\tb) =\nonumber\\& \sum_{P(2,3)} (T^{a_2}T^{a_3})_{i_1i_4} A^{(0)}_{4}(1_t,2,3,4_\tb) \\&
	\mc A^{(1)}_4(1_t,2,3,4_\tb) = \nonumber\\& \sum_{P(2,3)} N (T^{a_2}T^{a_3})_{i_1i_4} A^{(1)}_{4;1}(1_t,2,3,4_\tb)
	\nonumber\\&
	+\delta^{a_2a_3}\delta_{i_1i_4} A^{(1)}_{4;3}(1_t,4_\tb;2,3),
\end{align}
\begin{align}
	&A^{(1)}_{4;1}(1_t,2,3,4_\tb) = A^{[L]}(1_t,2,3,4_\tb)\nonumber\\&
	-\frac{1}{N^2}A^{[R]}(1_t,2,3,4_\tb-\frac{N_f}{N} A^{[f]}(1_t,2,3,4_\tb)\nonumber\\&
	-\frac{N_H}{N} A^{[H]}(1_t,2,3,4_\tb),
	\nonumber\\
	&A^{(1)}_{4;3}(1_t,4_\tb;2,3) = \sum_{P(2,3)} \bigg\{
		A^{[L]}(1_t,2,3,4_\tb)\nonumber\\&
		+A^{[L]}(1_t,2,4_\tb,3)
		+A^{[R]}(1_t,2,3,4_\tb)\bigg\}.
\end{align}
For the sake of brevity we restrict ourselves to the $++++$ helicity configuration. The remaining independent helicity
amplitude has been generated and checked against the literature but will be presented elsewhere. The tree level
amplitude can be written as,
\begin{align}
	&A^{(0)}_4(1_t^+,2^+,3^+,4_\tb^+) =\nonumber\\&im^3
	\frac{\B{2}{3}\A{\eta_1}{\eta_4}}{\A{2}{3}\AB{2}{1}{2}\A{\eta_1}{\fl 1}\A{\eta_4}{\fl 4}}.
\end{align}
We find it convenient to write down the one-loop amplitudes in terms of the scalar integrals (using the notation of
\cite{Ellis:2007qk}) together with two finite integral functions $F_4$ and $F_2$. By defining,
\begin{align}
	F_4&\left(0,m^2,m^2,0,s_{12},s_{23},m^2,m^2,0,m^2\right) =\nonumber\\&
		I_4\left(0,m^2,m^2,0,s_{12},s_{23},m^2,m^2,0,m^2\right) \nonumber\\&
		-\frac{1}{\AB 212}I_3\left(s_{23},m^2,m^2,0,m^2,m^2\right),
\end{align}
we absorb all the divergent $\tfrac{1}{\e}\log\big(\tfrac{\beta+1}{\beta-1}\big)$, with $\beta=\sqrt{1-\tfrac{4m^2}{s_{12}}}$, into the three-mass 
triangle coefficient. We can then make use of the IR constraint on the integral coefficients to simplify the result:
\begin{align}
  &C^{[R]}_{4;1|2|3|4}+\AB212 C^{[R]}_{3;2|3|41} =\nonumber\\& (s_{23}-2m^2)\AB212 A^{(0)}(1_t,2,3,4_{\tb}).
\end{align}
We also move the $\log(m^2)$ dependence into the on-shell bubble contributions by defining,
\begin{align}
		F_2&\left( s_{12},0,m^2 \right) =\nonumber\\& I_2\left( s_{12},0,m^2
		\right)-I_2\left(m^2,0,m^2\right),\\
		F_2&\left( s_{23},m^2,m^2 \right) =\nonumber\\& I_2\left( s_{23},m^2,m^2 \right)-I_2\left(0,m^2,m^2\right).
\end{align}
The remaining integrals have been computed analytically long ago \cite{Nason:1987xz,Beenakker:1988bq}. We have made use of {\tt qcdloop} 
\cite{Ellis:2007qk} in making numerical comparisons with the literature. The analytic expressions for the leading colour
primitive amplitudes are given in equations eqs.(\ref{eq:A1L}-\ref{eq:A1H}).

\begin{figure*}
\begin{align}
	&-iA^{[L]}(1_t^+,2^+,3^+,4_\tb^+) =
-\frac{ \la\eta_1\eta_4\ra[32]^2m^3}{\la\eta_1\fl{1}\ra\la\eta_4\fl{4}\ra}I_4\left(m^2,0,0,m^2,s_{12},s_{23},m^2,0,0,0\right)\nonumber\\&
-\frac{\left(2s_{12}\la\eta_1\eta_4\ra-\la\eta_1|(1+2)(2+3)|\eta_4\ra\right)[32]m^3}{\la\eta_1\fl{1}\ra\la\eta_4\fl{4}\ra\la23\ra\AB{2}{1}{2}^2}
F_2\left(s_{12},0,m^2\right)\nonumber\\&
	-\frac{i}{2}A^{(0)}_4(1_t^+,2^+,3^+,4_\tb^+)\left(I_2(m^2,0,m^2)-1\right)\nonumber\\&
	-\frac{m(\la \eta_1|(1+2)(2+3)|\eta_4\ra +\la\eta_1\eta_4\ra  \la 2|1|2]) [32]}
   	{2 \la \eta_1\fl 1\ra  \la\eta_4\fl 4\ra  \la 23\ra  \la 2|1|2]}
	-\frac{m(\la\eta_1\eta_4\ra  \la 2|1|2]+\la 2\eta_1\ra  \la 3\eta_4\ra [32])}
	{3 \la \eta_1\fl 1\ra  \la \eta_4\fl 4\ra  \la 23\ra ^2}, 
	\label{eq:A1L}
\end{align}
\end{figure*}
\begin{figure*}
\begin{align}
	&iA^{[R]}(1_t^+,2^+,3^+,4_\tb^+) = 
	-F_4\left(m^2,0,0,m^2,s_{12},s_{23},0,m^2,m^2,m^2\right)\nonumber\\&\times\Bigg(
	-\frac{\la2\eta_1\ra\la2\eta_4\ra\left(2m^2+\AB{2}{1}{2}\right)[32]^2m^3}{2\la\eta_1\fl{1}\ra\la\eta_4\fl{4}\ra\la23\ra\AB{2}{1}{3}}
	+\frac{\la3\eta_1\ra\la3\eta_4\ra\left(2m^2+\AB{2}{1}{2}\right)[32]^2m^3}{2\la\eta_1\fl{1}\ra\la\eta_4\fl{4}\ra\la23\ra\AB{3}{1}{2}}\nonumber\\&
	-\frac{\left(2\left(2m^2-s_{23}\right)\la\eta_1\eta_4\ra+2\la\eta_1|(1+2)(2+3)|\eta_4\ra\right)[32]m^3}{2\la\eta_1\fl{1}\ra\la\eta_4\fl{4}\ra\la23\ra}
	\Bigg)
	\nonumber\\&
	-I_3\left(s_{23},m^2,m^2,m^2,m^2,0\right)
	\frac{\left(2m^2-s_{23}\right)\la\eta_1\eta_4\ra[32]m^3}{\la\eta_1\fl{1}\ra\la\eta_4\fl{4}\ra\la23\ra\AB{2}{1}{2}}
	-I_3\left(s_{12},0,m^2,0,m^2,m^2\right)\nonumber\\&\times\Bigg(
	\frac{(2\la\eta_1\eta_4\ra\la23\ra+4\la2\eta_4\ra\la3\eta_1\ra)[32]m^3}{\la\eta_1\fl{1}\ra\la\eta_4\fl{4}\ra\la23\ra^2}
	+\frac{\la3\eta_1\ra\la3\eta_4\ra\AB{2}{1}{3}[32]m^3}{\la\eta_1\fl{1}\ra\la\eta_4\fl{4}\ra\la23\ra^2\AB{2}{1}{2}}
	-\frac{\la2\eta_1\ra\la2\eta_4\ra\AB{3}{1}{2}[32]m^3}{\la\eta_1\fl{1}\ra\la\eta_4\fl{4}\ra\la23\ra^2\AB{2}{1}{2}}\nonumber\\&
	+\frac{\la2\eta_1\ra\la2\eta_4\ra\AB{2}{1}{2}[32]m^3}{\la\eta_1\fl{1}\ra\la\eta_4\fl{4}\ra\la23\ra^2\AB{2}{1}{3}}
	-\frac{\la3\eta_1\ra\la3\eta_4\ra\AB{2}{1}{2}[32]m^3}{\la\eta_1\fl{1}\ra\la\eta_4\fl{4}\ra\la23\ra^2\AB{3}{1}{2}}
	\Bigg)
	-I_3\left(s_{23},0,0,m^2,m^2,m^2\right)\nonumber\\&\times\Bigg(
	\frac{\la2\eta_1\ra\la2\eta_4\ra[32]^2m^3}{2\la\eta_1\fl{1}\ra\la\eta_4\fl{4}\ra\la23\ra\AB{2}{1}{3}}
	-\frac{\la3\eta_1\ra\la3\eta_4\ra[32]^2m^3}{2\la\eta_1\fl{1}\ra\la\eta_4\fl{4}\ra\la23\ra\AB{3}{1}{2}}
	-\frac{\la\eta_1\eta_4\ra[32]m^3}{\la\eta_1\fl{1}\ra\la\eta_4\fl{4}\ra\la23\ra}
	\Bigg)\nonumber\\&
	-F_2\left(s_{12},0,m^2\right)
	\frac{\left(2s_{12}\la\eta_1\eta_4\ra-\la\eta_1|(1+2)(2+3)|\eta_4\ra\right)[32]m^3}{\la\eta_1\fl{1}\ra\la\eta_4\fl{4}\ra\la23\ra\AB{2}{1}{2}^2}
	\nonumber\\&
	+\frac{i}{2}A^{(0)}_4(1_t^+,2^+,3^+,4_\tb^+)\left(I_2(m^2,0,m^2)-1\right)
	+\frac{m(\la \eta_1|(1+2)(2+3)|\eta_4\ra +\la\eta_1\eta_4\ra  \la 2|1|2]) [32]}
   	{2 \la \eta_1\fl 1\ra  \la\eta_4\fl 4\ra  \la 23\ra  \la 2|1|2]},
	\label{eq:A1R}
\end{align}
\end{figure*}
\begin{figure*}
\begin{align}
	&-iA^{[H]}(1_t^+,2^+,3^+,4_\tb^+) = 
	-\frac{2 m(\la\eta_1\eta_4\ra\AB{2}{1}{2}+\la2\eta_1\ra\la3\eta_4\ra[32])}{\la\eta_1\fl{1}\ra\la\eta_4\fl{4}\ra\la23\ra^3[32]}
\nonumber\\&
\times\Bigg(
s_{23} m_H^2I_3\left(s_{23},0,0,m_H^2,m_H^2,m_H^2\right)
+2 m_H^2F_2\left(s_{23},m_H^2,m_H^2\right)
+\frac{1}{6}s_{23}
\Bigg).
\label{eq:A1H}
\end{align}
\end{figure*}

\section{Tadpole coefficients and IR/UV constraints}

The universal IR and UV behaviour provide strong constraints on the form of the terms proportional to $\log(m^2)$.
It has been proposed long ago by Bern and Morgan that this information can be used to completely fix the on-shell bubble
and tadpole coefficients \cite{Bern:1995db}. Here we show, for the $++++$ example considered above, how this can be
extended to deal with massive external particles. It is hoped that proceeding along similar lines would give a general
method for any massive amplitude. Using the universal pole structure \cite{Catani:2000ef,Mitov:2006xs} 
the overall coefficient of $1/\e-\log(m^2)$ can be described by:
\begin{align}
  &A^{(1)}(1_t,2,3,4_\tb) = \nonumber\\&
  A^{(1),cc}(1_t,2,3,4_\tb)+\mathcal{R}(1_t,2,3,4_\tb)\nonumber\\&
  -A^{(0)}(1_t,2,3,4_\tb)\frac{C_F}{\e(1-2\e)}\left( \frac{\mu_R^2}{m^2} \right)^\e.
\end{align}
In our example we find:
\begin{align}
  &C_{2|m^2}(1_t^+,2^+,3^+,4_\tb^+) =\nonumber\\& -2C_F C_{2;12}(1_t^+,2^+,3^+,4_\tb^+),\\&
  C_{1}(1_t^+,2^+,3^+,4_\tb^+) = \nonumber\\&-C_F A^{(0)}(1_t^+,2^+,3^+,4_\tb^+).
\end{align}
One can then find agreement with the $D$-dimensional rational term as defined in reference \cite{Ellis:2008ir} using:
\begin{align}
	&R^{DD,[X]}(1_t^+,2^+,3^+,4_\tb^+) = \nonumber\\&\mathcal{R}(1_t^+,2^+,3^+,4_\tb^+)+\frac{i}{2}A^{(0)}(1_t^+,2^+,3^+,4_\tb^+),
\end{align}
where $\mathcal{R}$ is defined according to eq. (\ref{eq:a1basis2}). The equation applies to both Left- and Right-moving
primitive amplitudes. We note that this simple relation holds only for the $++++$ helicity amplitude since the
``on-shell" bubble happens to cancel the ``off-shell" bubble precisely. For the general case a more complicated relation
would need to be determined.

\section{Outlook}

We have shown that combinations of generalised unitarity and optimised Feynman diagram computations
can be an efficient method to generate compact analytic expressions for massive one-loop amplitudes.
We have shown complete helicity amplitudes for $t\tb$ production via gluon fusion in the ``all-plus" configuration.
We have checked our results against those in the literature and found complete agreement in all cases. We hope that the
result presented here will allow for a faster evaluation of the NLO cross-section.

The methods employed are quite general and have also been successful in studying the more complicated helicity
configurations. However the full results for this process, including  the sub-leading colour contributions, will be
presented in a forthcoming publication\footnote{The full amplitudes have been subsequently presented in {\tt arXiv:1101.5947 [hep-ph]}}.\\

\noindent\textbf{Acknowledgements}\\

\noindent We would like to thank Keith Ellis, Sven Moch and Peter Uwer for valuable discussions.
This work was supported in part by the Helmholtz Gemeinschaft under contracts VH-NG-105, by the DFG within SFB/TR 9 
and Graduiertenkolleg GRK 1504, and by the ``HEPTOOLS" Training Network MRTN-CT-2006-035505.

\bibliographystyle{utphysmod}
\bibliography{ll10}

\end{document}